\documentclass[12pt]{article}
\usepackage{amsmath,amssymb,setspace,natbib}
\newtheorem{theorem}{Theorem}

\begin{document}
\title{Shrinkage Tuning Parameter Selection in Precision Matrices Estimation}         
\author{Heng Lian\\Division of Mathematical Sciences\\School of Physical and Mathematical Sciences\\Nanyang Technological University\\Singapore 637371}        
\date{}          
\maketitle

\section*{ABSTRACT}
Recent literature provides many computational and modeling approaches for covariance matrices estimation in a penalized Gaussian graphical models but relatively little study has been carried out on the choice of the tuning parameter. This paper tries to fill this gap by focusing on the problem of shrinkage parameter selection when estimating sparse precision matrices using the penalized likelihood approach. Previous approaches typically used K-fold cross-validation in this regard. In this paper, we first derived the generalized approximate cross-validation for tuning parameter selection which is not only a more computationally efficient alternative, but also achieves smaller error rate for model fitting compared to  leave-one-out cross-validation. For consistency in the selection of nonzero entries in the precision matrix, we employ a Bayesian information criterion which provably can identify the nonzero conditional correlations in the Gaussian model. Our simulations demonstrate the general superiority of the two proposed selectors in comparison with leave-one-out cross-validation, ten-fold cross-validation and Akaike information criterion.\\

\noindent\textbf{Keywords}: Adaptive Lasso; BIC; Generalized approximate cross-validation; Precision matrix; SCAD penalty. 

\section{INTRODUCTION}       
Undirected graphical models provide an easily understood way of describing and visualizing the relationships between multiple random variables. Usually the goal is to seek the simplest model that adequately explains the observations. In the Gaussian case, the zero entries in the inverse of the covariance matrix, or the precision matrix, correspond to independence of those two random variables conditional on all others. 

Efficient estimation of a covariance matrix or its inverse is an important statistical problem. In network modeling, for example, correct identification of the nonzero entries provides an understanding of the relationships between the expression levels of different genes. There also exists an abundance of methods in multivariate analysis that requires the estimation of the covariance matrix or its inverse, such as principal component analysis (PCA) or linear discriminant analysis (LDA). 

Due to the recent surge in interests in the estimation of the covariance matrix with the appearance of a large amount of data generated by modern experimental advances such as microarrays, a large number of different estimators have been proposed by now. Mostly motivated by the fact that the sample covariance matrix is typically a noisy estimator when the sample size is not significant and the resulting dense matrix is difficult to interpret, almost all modern estimators regularize the matrix by making it sparse. This notion of sparsity in the context of estimating covariance matrices has been noticed by some author as early as in \cite{dempster72} who simplified the matrix structure by setting some entries to zero. 

Traditionally, for the identification of zero elements, forward-selection or backward-selection is performed with each element tested at each step. However, it is computationally infeasible even for data with a moderate number of variables. \cite{ligui06} proposed using threshold gradient descent for estimating the sparse precision matrix in the Gaussian graphical models. \cite{bickel08} developed asymptotic theories on banding methods for both covariance and precision matrix estimation based on direct thresholding of the elements. Another way to estimate the graphical model is to perform a regression for each variable on all the remaining ones. For example, \cite{dobra04} used a Bayesian approach that employs a stochastic algorithm which can deal with tens of thousands of variables. 

Recently, there has been much interest on the estimation of sparse covariance or precision matrices using penalized likelihood method. \cite{meinshausen06} estimates the conditional independence separately for each random variable using the Lasso penalty. Note that setting up separate regressions for each node does not result in a valid likelihood for the covariance matrix and thus in order to obtain a positive-definite estimator, some post-processing is typically performed as the last step. \cite{yuan07} used semi-definite programming algorithms to achieve sparsity by penalizing the normal likelihood with Lasso penalty on the elements of the precision matrix.

The Lasso penalty will force some of the coefficients to be exactly zero. Compared to traditional model selection methods using information criteria, Lasso is continuous and thus more stable. 
However, several authors \citep{meinshausen06,zhaoyu06} have noted that Lasso is in general not consistent for model selection unless some nontrivial conditions on the covariates are satisfied. Even when those conditions are indeed satisfied, the efficiency of the estimator is compromised when one insists on variable selection consistency since the coefficients are over-shrunk. To address these shortcomings of Lasso, \cite{fan01} proposed the smoothly clipped absolute deviation (SCAD) penalty which takes into account several desired properties of the estimator such as sparsity, continuity, asymptotic unbiasedness. They also show that the resulting estimator possesses the oracle property, i.e. it is consistent for variable selection and behaves the same as  when the zero coefficients are known in advance. These results are extended to the case with a diverging number of covariates in \cite{fan04}. \cite{zou06} proposed adaptive Lasso  using a weighted $L_1$ penalty with weights determined by an initial estimator and similar oracle property followed. The idea behind the adaptive lasso is to assign higher penalty for zero coefficients and lower penalty for larger coefficients. 

In the context of Gaussian graphical models, \cite{lam07} studied rates of convergence of sparse covariance/precision matrices estimation via a general penalty function including SCAD and adaptive Lasso penalties as special cases and showed that these penalty functions attenuated the bias problem associated with Lasso penalty. In \cite{fan08}, through local linear approximation to the SCAD penalty function, efficient computation of the penalized likelihood is achieved using the graphical Lasso algorithm of \cite{friedman08}. Oracle properties as defined in \cite{fan01} were shown for the precision matrix estimator in \cite{fan08}.

The attractive oracle property of the penalized estimator depends critically on the appropriate choice of the tuning parameter. For penalized likelihood method, most of the studies mentioned above employed cross-validation (CV) for the selection of the tuning parameter. Cross-validation requires fitting the model based on different subsets of the observations multiple times, which increased the computational complexity of this approach. As an approximation to leave-one-out cross-validation (LOOCV), \cite{craven79} proposed generalized cross-validation (GCV) for smoothing spline, and further derived generalized approximate cross-validation (GACV) for non-Gaussian data \citep{dongwahba96}. We will follow similar strategies and derive GACV for the Gaussian graphical model that can be computed efficiently without multiple fitting of the model. However, for linear regression, \cite{wang07} showed that generalized cross validation cannot select the tuning parameter satisfactorily. In particular, the tuning parameter chosen by GCV tend to overestimate the model size. Because of the asymptotic equivalence of generalized cross-validation, leave-one-out cross-validation and Akaike information criterion (AIC) in this simple model, the authors proposed to use Bayesian information criterion (BIC) for consistent model selection. We will demonstrate that similar conclusions can be reached for our problem of precision matrices estimation.

In this paper, we estimate the precision matrices using the same computational approach as in \cite{fan08}. However, we focus on the problem of tuning parameter selection in penalized Gaussian graphical model. In the next section, two selectors, GACV and BIC, are proposed in this context. Simulation studies are carried out in Section 3 which demonstrate the superior performance of the proposed selectors. Finally, some remarks conclude the article in Section 4.

\section{PENALIZED ESTIMATION AND TUNING PARAMETER SELECTION}    
Suppose $\mathbf{x}_1,\ldots,\mathbf{x}_n$ are i.i.d. observations generated from a $p-$dimensional multivariate Gaussian distribution with mean $\mathbf{0}$ and unknown covariance matrix $\mathbf{\Sigma}_0$, where $\mathbf{x}_i=(x_{i1},\ldots,x_{ip})^T$. Denote the sample covariance matrix by $\mathbf{S}=\sum_{i=1}^n\mathbf{x}_i\mathbf{x}_i^T/n$. The inverse of $\mathbf{S}$ is a natural estimator of the precision matrix $\mathbf{\Omega}_0=\mathbf{\Sigma}^{-1}_0$ which is the estimator that maximizes the Gaussian log-likelihood
\[
\max_{\mathbf{\Omega}}\log|\mathbf{\Omega}|-\mbox{Tr}(\mathbf{S}\mathbf{\Omega}),
\]
where $|\mathbf{\Omega}|$ is the determinant of the matrix $\mathbf{\Omega}$.

However, if the true precision matrix is known to be sparse or the dimension of the random vector is large, the performance of the estimator can typically be improved by maximizing the penalized log-likelihood instead:
\[
\max_{\mathbf{\Omega}}\log|\mathbf{\Omega}|-\mbox{Tr}(\mathbf{S}\mathbf{\Omega})-\sum_{i=1}^p\sum_{j=1}^p p_{\lambda_{ij}}(|\omega_{ij}|),
\]
where $\omega_{ij}, i=1,\ldots,p,\, j=1,\ldots,p, $ are the entries of $\mathbf{\Omega}$ and $\lambda_{ij}$ are the tuning parameters, which for now are left unspecified to allow for very general penalty functions.
Note that we also penalize the diagonal elements as in \cite{fan08}.

In the above, using the penalty $\lambda_{ij}=\lambda, p_\lambda(|x|)=\lambda|x|$ produces the Lasso estimator, while using $\lambda_{ij}=\lambda/|\tilde{\omega}_{ij}|^\gamma$, where $\gamma>0$ and $\tilde{\omega}_{ij}$ is any consistent estimator of $\omega_{ij}$, paired with the same $p_\lambda(|x|)=\lambda|x|$ produces the adaptive Lasso estimator. Another commonly used penalty function proposed by \cite{fan01} is the SCAD penalty defined by its derivative
\[
p_{\lambda}(|x|)=\lambda I(|x|\le\lambda)+\frac{(a\lambda-|x|)_+}{(a-1)}I(|x|>\lambda),
\]
with $a=3.7$ according to the suggestion made in \cite{fan01}. 
Unlike the Lasso estimator, which produces substantial biases for large elements, the adaptive Lasso as well as the SCAD estimator achieved the oracle property as defined in \cite{fan01}, which estimates the precision matrix as well as in the ideal situation where the zero elements are known. Efficient maximization with either Lasso or adaptive Lasso penalty can be directly performed using the graphical Lasso algorithm \citep{friedman08}. To take advantage of graphical Lasso algorithm, \cite{fan08} suggested using local linear approximation to recast the SCAD penalized likelihood as weighted Lasso in each step. It was pointed out that a one-step algorithm can perform as well as the fully iterative local linear approximation algorithm. The reader are referred to \cite{fan08} for further details.

Here we focus on tuning parameter selection which consists of a single parameter $\lambda$ for all three penalty functions mentioned above. In \cite{fan08}, it was shown that for both adaptive Lasso penalty and the SCAD penalty, when $\lambda\rightarrow 0$ and $\sqrt{n}\lambda\rightarrow\infty$, the resulting estimators attain the oracle property. Hence, the choice of $\lambda$ is critical. In \cite{fan08}, they proposed to use K-fold cross-validation (KCV), with $K=10$ probably the most popular choice in the literature. In K-fold cross-validation, one partitions the data into $K$ disjoint subsets and chooses $\lambda$ that maximizes the score
\[
KCV(\lambda)=\sum_{k=1}^Kn_k\left(\log|\hat{\mathbf{\Omega}}^{(-k)}(\lambda)|-\mbox{Tr}(\mathbf{S}^{(-k)}\hat{\mathbf{\Omega}}^{(-k)}(\lambda))\right),
\]
where $n_k$ is the sample size of the $k-$th partition, $\mathbf{S}^{(-k)}$ is the sample covariance matrix based on the data with $k-$th partition excluded, and  $\hat{\mathbf{\Omega}}^{(-k)}(\lambda)$ is the estimate of the precision matrix based on the data excluding the $k-$th partition with $\lambda$ as the tuning parameter. 

The usual leave-one-out cross-validation (LOOCV) is just a special case of the above with $K=n$ so that each partition consists of one single observation. Computation of KCV entails $K$ maximization problems fitted with each partition deleted in turn for each fixed $\lambda$ as well as a final maximization using the optimal tuning parameter. Thus the computation of KCV is infeasible for large datasets especially when $K$ is large. In the smoothing spline literature, the most popular approach for tuning parameter selection is the generalized cross-validation (GCV) for Gaussian data. For non-Gaussian data, \cite{dongwahba96} proposed the generalized approximate cross-validation (GACV), which is obtained by constructing an approximation to the LOOCV based on the log-likelihood. The formula presented there does not directly apply to our problem since their derivation is based on regression problems. We also derive a GACV score based on an approximation to the LOOCV for our problem. The derivation of GACV is complicated by the multivariate nature of each left-out observation. The detail is deferred to Appendix A. Maybe surpringsingly, even though GACV is motivated by an efficient approximation to LOOCV, it almost always performs better than LOOCV and sometimes better than ten-fold CV as demonstrated by the simulation studies in the next section.
 
In classical model selection literature, it is well understood that CV, GCV and AIC share similar asymptotic properties. All these criteria are loss efficient but inconsistent for model selection \citep{shao97,yang05}. For penalized linear regression with the SCAD penalty, it has been verified by \cite{wang07} that GCV is not able to identify the true model consistently when it is used for tuning parameter selection. To address this problem, \cite{wang07} employed a variable selection criterion known to be consistent in the classical literature, BIC, as the tuning parameter selector and proved that the resulting tuning parameter can identify the true model consistently. Similar conclusion has been drawn for adaptive Lasso \citep{wangleng07}. In \cite{wang08}, the theory is further extended to linear regression problems with a diverging number of parameters. In our context, with BIC, we select the optimal $\lambda$ by minimizing
\[
BIC(\lambda)=-\log|\hat{\mathbf{\Omega}}(\lambda)|+\mbox{Tr}(\mathbf{S}\hat{\mathbf{\Omega}}(\lambda))+k\frac{\log n}{n},
\]
where $k$ is the number of nonzero entries in the upper diagonal portion of the estimated precision matrix $\hat{\mathbf{\Omega}}(\lambda)$.

We conjecture that our GACV proposed above is also not appropriate for model selection, although a formal proof seems illusive due to the complicated form of the GACV score. Nevertheless, we can extend the consistency proof for BIC using empirical process theory to show that the tuning parameter selected by BIC will result in consistent model identification. The result is stated below while the proof is relegated to Appendix B.
\begin{theorem}
Denoting the optimal tuning parameter selected using BIC by $\hat{\lambda}_{BIC}$, if Conditions  1-3 in Appendix B hold, then the penalized likelihood estimator correctly identifies all the zero elements in the true precision matrix. That is,  $pr(\mathcal{S}_{\hat{\lambda}_{BIC}}=\mathcal{S}_T)\rightarrow 1$, where $\mathcal{S}_{\lambda}=\{(i,j): i\le j, \hat{\omega}_{ij}(\lambda)\neq 0\}$ is the set of nonzero entries above and including the diagonal in the estimated precision matrix and $\mathcal{S}_T$ is similarly defined to be the set of nonzero entries in the true precision matrix $\mathbf{\Omega}_0$.
\end{theorem}
The conditions imposed on the penalty function can be verified for both SCAD penalty and adaptive Lasso penalty. Thus for these two penalty functions, BIC identifies the correct model for the precision matrix.

\section{SIMULATIONS}
In this section we compare the performance of different tuning parameter selectors for Lasso, adaptive Lasso and SCAD penalty estimators in Gaussian graphical models. For adaptive Lasso penalty with $\lambda_{ij}=\lambda/|\tilde{\omega}_{ij}|^\gamma$, we use the Lasso estimator as the initial consistent estimator and set $\gamma=0.5$ following \cite{fan08}.   Besides CV, KCV (ten-fold), GACV and BIC, we also use AIC as the tuning parameter selector, which is defined by 
\[AIC(\lambda)=-\log|\hat{\mathbf{\Omega}}(\lambda)|+\mbox{Tr}(\mathbf{S}\hat{\mathbf{\Omega}}(\lambda))+2k/n , \]
where $k$ is the number of nonzero entries in the upper diagonal part of $\hat{\mathbf{\Omega}}(\lambda)$.

 We use three examples with different covariance structures in our simulation. The first one has a tridiagonal precision matrix:
\[\mathbf{\Omega}_0: \omega^0_{ii}=1, \omega^0_{i,i-1}=\omega^0_{i,i+1}=0.5.\]
The second one has a dense precision matrix with exponential decay:
\[\mathbf{\Omega}_0: \omega^0_{ij}=0.5^{|i-j|},\]
where no entries are exactly zero but many are so small that penalization is expected to reduce the variability of the estimators.
The third one has a more general sparse precision matrix. For each non-diagonal element $\omega^0_{ij}, i<j$ of $\mathbf{\Omega}_0$, with probability $0.8$ we set it to be zero, otherwise we sample a value for $\omega_{ij}^0$ from a uniform distribution on $[-1,-0.5]\cup[0.5,1]$. Each diagonal entry is set as twice of the sum of the absolute values of the corresponding row elements excluding the diagonal entry itself.

 For each example, we use $n$ i.i.d. random vectors generated from a multivariate Gaussian distribution $N(0,\mathbf{\Omega}_0^{-1})$ with dimension $p=20$. We consider both $n=50$ and $n=100$. The errors are calculated based on 500 simulated datasets in each example. To compare the performance of different selectors, we use the Frobenius norm $||\mathbf{\Omega}_0-\hat{\mathbf{\Omega}}||_F$ as well as the entropy loss \citep{yuan07} defined by 
\[ \mbox{Entropy}(\mathbf{\Omega}_0,\hat{\mathbf{\Omega}})=-\log|\mathbf{\Omega}_0^{-1}\hat{\mathbf{\Omega}}|+\mbox{Tr}(\mathbf{\Omega}_0^{-1}\hat{\mathbf{\Omega}})-p .\]
For the performance in terms of sparsity of the matrix, we use false positives (number of zero entries in the true precision matrix identified to be nonzero) and false negatives (number of nonzero entries in the true precision matrix identified to be zero).

The results for the Gaussian models are summarized in Tables 1-3 for Lasso penalty, adaptive Lasso penalty and SCAD penalty respectively, which gives the average losses and the average number of false positives and negatives for each case together with the corresponding standard errors. For sparse precision matrices, the BIC approach outperforms LOOCV, KCV and AIC in terms of the two loss measures as well as the sum of false positives and false negatives. For dense matrices, although the number of false negatives is generally larger compared to other selectors, which is certainly as expected, the performance of BIC in terms of loss is still superior. Based on our simulations, tuning parameter selected by AIC generally performs the worst. Finally, maybe surprisingly, the performance of GACV is almost always better than LOOCV, and in many cases also better than KCV. The reader should note that GACV can be computed much faster than either LOOCV or ten-fold CV.

\begin{table}
{\scriptsize\begin{tabular}{ccccccc}
&&BIC&AIC&CV&KCV&GACV\\
Tridiagonal &&&&&\\
n=100&Frobenius&7.02&17.40&7.11&7.21&6.37\\
&&(1.44)&(7.60)&(1.24)&(1.38)&(1.12)\\
&Entropy&0.80&1.85&0.88&0.89&0.90\\
&&(0.17)&(0.51)&(0.14)&(0.15)&(0.15)\\
&FP&98.36&205.97&164.93&153.17&149.61\\
&&(35.46)&(32.49)&(16.80)&(19.09)&(32.08)\\
&FN&2.63&0.16&2.90&3.06&0.98\\
&&(2.60)&(0.55)&(2.60)&(2.67)&(1.34)\\
n=50&Frobenius&11.39&70.03&12.36&12.35&12.03\\
&&(2.93)&(35.30)&(2.02)&(1.99)&(5.68)\\
&Entropy&1.38&7.29&1.41&1.47&1.45\\
&&(0.35)&(2.72)&(0.19)&(0.17)&(0.49)\\
&FP&109.51&212.61&190.37&170.42&171.36\\
&&(51.95)&(25.04)&(33.86)&(35.81)&(35.87)\\
&FN&7.84&0.77&6.15&6.21&4.53\\
&&(5.55)&(1.56)&(2.52)&(3.00)&(3.47)\\
Dense&&&&&&\\
n=100&Frobenius& 2.08&5.31&2.94&2.08&1.84\\
&&(0.50)&(2.36)&(0.38)&(0.54)&(0.24)\\
&Entropy&0.93&2.05&0.97&0.94&0.94\\
&&(0.17)&(0.5)&(0.12)&(0.17)&(0.16)\\
&FP&0&0&0&0&0\\
&&(0)&(0)&(0)&(0)&(0)\\
&FN&258.24&54.38&243.64&257.50&190.36\\
&&(51.63)&(30.45)&(37.94)&(42.15)&(39.54)\\
n=50&Frobenius& 3.32&27.97&3.57&4.21&3.68\\
&& (1.03)&(13.34)&(0.75)&(0.23)&(1.38)\\
&Entropy&1.48&6.58&1.56&1.75&1.55\\
&&(0.38)&(1.79)&(0.26)&(0.15)&(0.54)\\
&FP&0&0&0&0&0\\
&&(0)&(0)&(0)&(0)&(0)\\
&FN&256.57&34.42&224.57&256.85&170.43\\
&&(76.06)&(32.89)&(42.33)&(9.63)&(44.12)\\
Random Sparse&&&&&&\\
n=100&Frobenius& 25.52& 52.63&27.39&26.19&22.69\\
&&(8.84)& (29.36)&(7.24)&(8.32)&(6.42)\\
&Entropy&0.87&1.89&0.91&0.89&0.87\\
&&(0.1)&(0.53)&(0.11)&(0.12)&(0.13)\\
&FP&60.56&156.34&91.53&83.16&131.10\\
&&(31.16)&(27.77)&(27.11)&(23.16)&(37.72)\\
&FN&47.35&8.40&44.59&45.34&30.81\\
&&(13.74)&(6.23)&(11.55)&(12.61)&(12.21)\\
n=50&Frobenius& 33.42&93.59&34.41&34.69&34.74\\
&& (17.80)&(46.31)&(9.68)&(10.24)&(17.53)\\
&Entropy&1.36&5.38&1.52&1.51&1.55\\
&&(0.49)&(2.27)&(0.17)&(0.17)&(0.56)\\
&FP&73.19&277.11&95.40&94.89&139.02\\
&&(48.83)&(27.08)&(18.37)&(19.13)&(50.53)\\
&FN&58.17&6.68&54.89&55.36&35.83\\
&&(14.66)&(5.85)&(11.00)&(11.19)&(13.43)\\
\end{tabular}}
\caption{{\small Simulation results for Lasso estimators using five different tuning parameter selectors. The reported average errors are based on 500 simulated datasets. The numbers in the brackets are the corresponding standard errors. }}
\end{table}

\begin{table}
{\scriptsize\begin{tabular}{ccccccc}
&&BIC&AIC&CV&KCV&GACV\\
Tridiagonal&&&&&\\
n=100&Frobenius&5.74&22.49&7.24&7.11&6.31\\
&&(1.64)&(10.01)&(2.38)&(2.21)&(2.40)\\
&Entropy&0.78&1.88&0.91&0.90&0.82\\
&&(0.18)&(0.57)&(0.23)&(0.21)&(0.24)\\
&FP&58.08&197.77&94.63&92.63&73.01\\
&&(22.60)&(55.73)&(15.33)&(14.26)&(23.06)\\
&FN&2.54&0.41&0.82&0.87&1.45\\
&&(2.79)&(0.88)&(1.19)&(1.25)&(1.77)\\
n=50&Frobenius&13.45&95.56&20.21&17.97&13.94\\
&&(6.81)&(47.25)&(9.08)&(7.11)&(6.06)\\
&Entropy&1.65&5.42&2.13&1.99&1.69\\
&&(0.49)&(1.64)&(0.58)&(0.47)&(0.44)\\
&FP&65.21&277.63&104.52&99.21&71.52\\
&&(26.98)&(57.21)&(20.21)&(19.15)&(26.92)\\
&FN&7.15&2.21&4.42&4.68&6.73\\
&&(5.33)&(3.07)&(3.39)&(3.58)&(5.18)\\
Dense&&&&&\\
n=100&Frobenius&1.63&6.25&2.14&1.91&1.75\\
&&(0.64)&(2.76)&(0.62)&(0.52)&(0.44)\\
&Entropy&0.79&1.98&0.99&0.92&0.85\\
&&(0.17)&(0.57)&(0.20)&(0.18)&(0.17)\\
&FP&0&0&0&0&0\\
&&(0)&(0)&(0)&(0)&(0)\\
&FN&285.72&133.80&233.08&243.84&266.06\\
&&(29.37)&(54.91)&20.58)&(19.12)&(28.14)\\
n=50&Frobenius& 4.29&29.63&5.34&5.27&4.04\\
&& (3.00)&(14.67)&(1.56)&(1.68)&(1.85)\\
&Entropy&1.84&5.94&2.26&2.23&1.82\\
&&(0.78)&(1.54)&(0.37)&(0.42)&(0.54)\\
&FP&0&0&0&0&0\\
&&(0)&(0)&(0)&(0)&(0)\\
&FN&267.14&106.85&226.00&227.57&264.71\\
&&(41.44)&(49.48)&(16.11)&(19.02)&(40.81)\\
General sparse&&&&&\\
n=100&Frobenius&19.35&55.67&25.06&23.87&20.13\\
&&(5.37)&(26.73)&(9.55)&(8.56)&(6.29)\\
&Entropy&0.77&1.88&1.03&0.99&0.82\\
&&(0.13)&(0.45)&(0.21)&(0.20)&0.17)\\
&FP&29.38&215.02&85.22&79.18&44.22\\
&&(19.88)&(45.96)&(18.72)&(21.08)&(24.29)\\
&FN&54.65&17.20&44.26&45.43&51.36\\
&&(11.53)&(11.42)&(8.62)&(9.28)&(10.96)\\
n=50&Frobenius&34.35&105.55&50.31&46.18&39.06\\
&&(19.80)&(59.75)&(14.94)&(19.53)&(31.00)\\
&Entropy&1.40&4.61&1.97&1.81&1.54\\
&&(0.48)&(1.33)&(0.36)&(0.46)&(0.69)\\
&FP&49.11&212.26&94.42&83.44&58.26\\
&&(28.73)&(47.45)&(15.51)&(25.35)&(38.58)\\
&FN&54.34&21.53&45.11&47.91&52.72\\
&&(11.19)&(10.21)&(8.82)&(11.06)&(11.86)\\
\end{tabular}}
\caption{{\small Simulation results for adaptive Lasso estimators using five different tuning parameter selectors.  }}
\end{table}

\begin{table}
{\scriptsize\begin{tabular}{ccccccc}
&&BIC&AIC&CV&KCV&GACV\\
Tridiagonal&&&&&\\
n=100&Frobenius&5.84&22.80&10.29&8.02&6.07\\
&&(4.06)&(7.20)&(3.29)&(2.83)&(5.13)\\
&Entropy&0.90&2.20&1.25&1.06&1.10\\
&&(0.26)&(0.44)&(0.32)&(0.31)&(0.41)\\
&FP&64.08&158.79&99.45&61.23&84.93\\
&&(20.58)&(26.89)&(22.97)&(20.59)&(26.51)\\
&FN&3.78&1.32&3.11&3.28&3.31\\
&&(2.93)&(1.38)&(2.94)&(1.58)&(2.54)\\
n=50&Frobenius&12.39&62.83&18.95&15.38&18.39\\
&&(8.76)&(36.70)&(11.29)&(4.24)&(11.22)\\
&Entropy&1.71&6.55&2.18&1.94&2.17\\
&&(1.41)&(1.71)&(0.76)&(0.38)&(1.74)\\
&FP&65.48&180.72&97.79&81.38&89.23\\
&&(23.53)&(27.83)&(23.24)&(15.75)&(25.80)\\
&FN&8.15&4.76&4.85&7.69&7.84\\
&&(4.45)&(2.85)&(5.66)&(7.16)&(4.21)\\
Dense&&&&&\\
n=100&Frobenius&1.68&6.75&1.94&1.92&1.72\\
&&(0.78)&(2.32)&(0.23)&(0.23)&(1.25)\\
&Entropy&0.82&2.37&0.91&0.89&0.89\\
&&(0.20)&(0.54)&(0.12)&(0.12)&(0.23)\\
&FP&0&0&0&0&0\\
&&(0)&(0)&(0)&(0)&(0)\\
&FN&268.78&112.72&203.94&202.06&212.92\\
&&(51.99)&(30.07)&(31.79)&(21.59)&(46.41)\\
n=50&Frobenius& 2.64&29.53&3.04&3.34&3.11\\
&& (1.87)&(10.51)&(0.52)&(0.57)&(2.30)\\
&Entropy&1.23&7.03&1.48&1.57&1.37\\
&&(0.83)&(1.29)&(0.19)&(0.19)&(1.32)\\
&FP&0&0&0&0&0\\
&&(0)&(0)&(0)&(0)&(0)\\
&FN&244.80&103.87&219.14&237.14&221.14\\
&&(81.30)&(12.99)&(32.97)&(32.97)&(65.51)\\
General sparse&&&&&\\
n=100&Frobenius&19.99&72.61&30.87&21.63&20.68\\
&&(6.69)&(26.93)&(17.37)&(6.90)&(8.05)\\
&Entropy&0.78&2.41&1.26&0.92&0.95\\
&&(0.31)&(0.39)&(0.38)&(0.24)&(0.27)\\
&FP&73.42&155.89&109.65&81.25&84.29\\
&&27.41)&(24.76)&(21.94)&(17.53)&23.37)\\
&FN&58.10&26.51&49.84&50.07&41.29\\
&&(12.25)&(9.12)&(11.46)&(12.54)&(11.64)\\
n=50&Frobenius&34.58&133.92&45.50&39.37&39.85\\
&&(25.36)&(86.84)&(15.84)&(14.19)&(22.57)\\
&Entropy&1.06&5.43&2.04&1.24&1.31\\
&&(0.45)&(1.65)&(0.68)&(0.52)&(10.41)\\
&FP&67.75&167.04&102.08&83.32&79.14\\
&&(26.64)&(36.23)&(23.48)&(15.30)&(25.94)\\
&FN&53.85&29.95&38.51&44.34&44.00\\
&&(10.29)&(10.65)&(11.05)&(9.37)&(9.25)\\
\end{tabular}}
\caption{{\small Simulation results for SCAD estimators using five different tuning parameter selectors.  }}
\end{table}

\section{CONCLUDING REMARKS}
In this paper we compare several approaches for tuning parameter selection in penalized Gaussian graphical models. As an approximation to leave-one-out cross-validation, we  derived generalized approximate cross-validation in the current context which is much faster to compute. Simulations show that GACV even outperforms the leave-one-out version. For model identification, we employ BIC for tuning parameter selection, and proved its consistency property. In our simulations with sparse matrices or dense matrices with many small entries, tuning parameter selected based on BIC clearly has better performance than all other approaches. 
\section*{APPENDIXES}
\subsection*{A. Derivation of GACV}
Denote the log-likelihood function by 
\[L(\mathbf{S},\mathbf{\Omega})=\log|\mathbf{\Omega}|-\mbox{Tr}(\mathbf{S}\mathbf{\Omega}).\]
In this section only, to simplify notation, the shrinkage estimator $\hat{\mathbf{\Omega}}(\lambda)$ is simply denoted by $\mathbf{\Omega}$ and similarly $\hat{\mathbf{\Omega}}^{(-i)}(\lambda)$ is denoted by $\mathbf{\Omega}^{(-i)}$. Thus it is implicitly understood that the estimators depend on a fixed $\lambda$. Let  $\mathbf{X}_i=\mathbf{x}_i\mathbf{x}_i^T$ be the covariance matrix based on a single observation so that $\mathbf{S}=\sum_{i=1}^n \mathbf{X}_i/n$. The LOOCV score is defined by 
\begin{eqnarray*}
CV(\lambda)&=&\sum_{i=1}^n\log|\mathbf{\Omega}^{(-i)}|-\mbox{Tr}(\sum_i{\mathbf{X}_i}\mathbf{\Omega}^{(-i)})=\sum_{i=1}^n L(\mathbf{X}_i,\mathbf{\Omega}^{(-i)})\\
&=&nL(\mathbf{S},\mathbf{\Omega})+\sum_{i=1}^n \left(L(\mathbf{X}_i,\mathbf{\Omega}^{(-i)})-L(\mathbf{X}_i,\mathbf{\Omega})\right)\\
&\approx&nL(\mathbf{S},\mathbf{\Omega})+\sum_{i=1}^n\left[\frac{dL(\mathbf{X}_i,\mathbf{\Omega})}{d\mathbf{\Omega}}\right]^Td\mathbf{\Omega},
\end{eqnarray*}
where we interpret $dL(\mathbf{X}_i,\mathbf{\Omega})/d\mathbf{\Omega}=dL(\mathbf{X}_i,\mathbf{\Omega})/d\mbox{vec}(\mathbf{\Omega})$ to be a $p^2-$dimensional column vector of partial derivatives and similarly $d\mathbf{\Omega}=\mbox{vec}(\mathbf{\Omega}^{(-i)}-\mathbf{\Omega})$. Besides, here as well as in the following, like the definition of generalized degrees of freedom in \cite{fan01} and \cite{wang07}, the partial derivatives corresponding to the zero elements in $\mathbf{\Omega}$ are ignored.

Using matrix calculus such as presented in \cite{bishop06}, we have 
\[\frac{dL(\mathbf{X}_i,\mathbf{\Omega})}{d\mathbf{\Omega}}=\mbox{vec}(\mathbf{\Omega}^{-1}-\mathbf{X}_i)\]
and we only need to deal with the term $d\mathbf{\Omega}=\mbox{vec}(\mathbf{\Omega}^{(-i)}-\mathbf{\Omega})$.

Denote the penalized log-likelihood based on the sufficient statistic $\mathbf{S}$ by $\bar{L}(\mathbf{S},\mathbf{\Omega})=L(\mathbf{S},\mathbf{\Omega})-\sum_{i,j}p_{\lambda_{ij}}(|\omega_{ij}|)$, Taylor expansion gives us
\begin{eqnarray*}
0&=&\frac{d\bar{L}(\mathbf{S}^{(-i)},\mathbf{\Omega}^{(-i)})}{d\mathbf{\Omega}}\\
&\approx&\frac{d\bar{L}(\mathbf{S},\mathbf{\Omega})}{d\mathbf{\Omega}}+
	\frac{d^2\bar{L}(\mathbf{S},\mathbf{\Omega})}{d\mathbf{\Omega}^2}d\mathbf{\Omega}+
	\frac{d^2\bar{L}(\mathbf{S},\mathbf{\Omega})}{d\mathbf{\Omega} d\mathbf{S}}d\mathbf{S},
\end{eqnarray*}
where $d^2\bar{L}(\mathbf{S},\mathbf{\Omega})/d\mathbf{\Omega}^2=d(\bar{L}(\mathbf{S},\mathbf{\Omega})/d\mbox{vec}(\mathbf{\Omega}))/d\mbox{vec}(\mathbf{\Omega})$ is the $p^2\times p^2$ Hessian matrix, and $d^2\bar{L}(\mathbf{S},\mathbf{\Omega})/d\mathbf{\Omega} d\mathbf{S}$ is defined similarly. Like before, $d\mathbf{\Omega}=\mbox{vec}(\mathbf{\Omega}^{(-i)}-\mathbf{\Omega})$ and $d\mathbf{S}=\mbox{vec}(\mathbf{S}^{(-i)}-\mathbf{S})$ actually denote their vectorized version.

Since $d\bar{L}(\mathbf{S},\mathbf{\Omega})/d\mathbf{\Omega}=0$, it immediately follows that
\[d\mathbf{\Omega}=-(\frac{d^2\bar{L}(\mathbf{S},\mathbf{\Omega})}{d\mathbf{\Omega}^2})^{-1}\frac{d^2\bar{L}(\mathbf{S},\mathbf{\Omega})}{d\mathbf{\Omega} d\mathbf{S}}d\mathbf{S}.\]

From matrix calculus, we have $d^2\bar{L}(\mathbf{S},\mathbf{\Omega})/d\mathbf{\Omega} d\mathbf{S}=-\mathbf{I}_{p^2\times p^2}$ and $d^2\bar{L}(\mathbf{S},\mathbf{\Omega})/d\mathbf{\Omega}^2=-(\mathbf{\Omega}^{-1}\otimes\mathbf{\Omega}^{-1}+\mathbf{D})$ where $\mathbf{D}$ is a diagonal matrix with diagonal elements $-p_{\lambda_{ij}}''(|\omega_{ij}|)$. Thus we have the approximation $d\mathbf{\Omega}\approx \left[d^2\bar{L}(\mathbf{S},\mathbf{\Omega})/d\mathbf{\Omega}^2\right]^{-1}d\mathbf{S}$. Even for moderate $p$, inversion of this $p^2\times p^2$ matrix is computationally infeasible. However, note that typically we consider only the situation with $\lambda=o(1)$ and $p_{\lambda_{ij}}''(|\omega_{ij}|)=o(1)$ (for example, the second derivative for SCAD penalty function is exactly zero for nonzero elements). Thus we can approximate $(\mathbf{\Omega}^{-1}\otimes\mathbf{\Omega}^{-1}+\mathbf{D})^{-1}$ by $(\mathbf{\Omega}^{-1}\otimes\mathbf{\Omega}^{-1})^{-1}=\mathbf{\Omega}\otimes\mathbf{\Omega}$ and $d\mathbf{\Omega}\approx -(\mathbf{\Omega}\otimes\mathbf{\Omega})d\mathbf{S}=\mbox{vec}(\mathbf{\Omega}\cdot(\mathbf{S}^{(i)}-\mathbf{S})\cdot\mathbf{\Omega})$ which involves only $p\times p$ matrices and no inversion of matrices is required.

\subsection*{B. Proof of Theorem 1}
We only need to assume the general conditions on the penalty function that guarantees the oracle property of the estimator with appropriately chosen tuning parameter. In particular, we assume that 

\textit{Condition 1}. $\max\{|p_{\lambda_n}''(|\omega^0_{ij}|):\omega_{ij}^0\neq 0\}\rightarrow 0.$

\textit{Condition 2}. $\liminf_{n\rightarrow\infty}\liminf_{x\rightarrow 0^+} p_{\lambda_n}'(x)/\lambda_n>0$.

\textit{Condition 3}. The (theoretically) optimal tuning parameter satisfies $\lambda_n\rightarrow 0$ and $\sqrt{n}\lambda_n\rightarrow\infty$.

 For an arbitrary model $\mathcal{S}$ specified by the constraints that only some of the elements in the precision matrix can be nonzero, i.e. $\mathcal{S}\subseteq\{(i,j): i\le j\}$ is the set of elements not constrained to be zero, denote by $L_\mathcal{S}$ the value of the constrained maximized likelihood with infinite data: $L_\mathcal{S}=\max_\mathbf{\Omega} E(\log|\mathbf{\Omega}|-\mbox{Tr}(\mathbf{X}\mathbf{\Omega}))$, where the maximization is performed over $\mathbf{\Omega}$ with zero entries for all $(i,j)\in \mathcal{S}$.  We partition $\Lambda=[0,\infty)$ into three parts: $\Lambda_0=\{\lambda\in\Lambda: \mathcal{S}_\lambda=\mathcal{S}_T\}, \Lambda_{-}=\{\lambda\in\Lambda: \mathcal{S}_\lambda\nsupseteq \mathcal{S}_T\}, \Lambda_{+}=\{\lambda\in\Lambda: \mathcal{S}_\lambda\supsetneq \mathcal{S}_T\}$, where $\mathcal{S}_\lambda$ is the model identified by the estimator when $\lambda$ is used as the tuning parameter, and $\mathcal{S}_T$ is the true model $\mathcal{S}_T=\{(i,j): i\le j, \omega_{ij}^0\neq 0\}$. We will prove separately that under-fitting probability and over-fitting probability are both negligible.

\textit{Bounding the under-fitting probability}: If $\mathcal{S}_\lambda\nsupseteq \mathcal{S}_T$, we have that 
\begin{eqnarray*}
BIC(\lambda)&=&-L(\mathbf{S},\hat{\mathbf{\Omega}}(\lambda))+|\mathcal{S}_{\lambda}|\frac{\log n}{n}\\
&\ge& -L(\mathbf{S},\hat{\mathbf{\Omega}}_{\mathcal{S}_\lambda})+|\mathcal{S}_{\lambda}|\frac{\log n}{n}\\
&\stackrel{P}{\rightarrow}& -L_{\mathcal{S}_\lambda},
\end{eqnarray*}
where $\hat{\mathbf{\Omega}}_{\mathcal{S}}$ is the unpenalized maximum likelihood estimator based on model $\mathcal{S}$. The convergence above is based on the uniform convergence of the empirical distribution since the class of log-likelihood functions is Glivenko-Catelli. 

Similarly, with the optimal choice of $\lambda_n$ that satisfies Condition 3, 
\[
BIC(\lambda_n)=-L(\mathbf{S},\hat{\mathbf{\Omega}}(\lambda_n))+|\mathcal{S}_{\lambda_n}|\frac{\log n}{n}\stackrel{P}{\rightarrow} -L_{\mathcal{S}_T}.
\] 
Thus we have $\mbox{pr}\{\inf_{\lambda\in\Lambda_{-}}BIC(\lambda)>BIC(\lambda_n)\}\rightarrow 1$ since $-L_{\mathcal{S}_T}<-L_{\mathcal{S}}$ when $\mathcal{S}\nsupseteq \mathcal{S}_T$.

\textit{Bounding the over-fitting probability}: Now suppose $\lambda\in\Lambda_{+}$. Since 
\[BIC(\lambda_n)=-L(\mathbf{S},\hat{\mathbf{\Omega}}(\lambda_n))+|\mathcal{S}_{\lambda_n}|\frac{\log n}{n}\]
and 
\[BIC(\lambda)=-L(\mathbf{S},\hat{\mathbf{\Omega}}(\lambda))+|\mathcal{S}_{\lambda}|\frac{\log n}{n},\] 
with $|\mathcal{S}_{\lambda}|>|\mathcal{S}_{\lambda_n}|=|\mathcal{S}_T|$ (with probability $1$), we will get $\mbox{pr}(\inf_{\lambda\in\Lambda_{+}}BIC(\lambda)>BIC(\lambda_n))\rightarrow 1$ if it can be shown that $L(\mathbf{S},\hat{\mathbf{\Omega}}(\lambda))-L(\mathbf{S},\hat{\mathbf{\Omega}}(\lambda_n))=O_p(1/n)$. 

We will use the usual notion for sample mean: $P_nL(\mathbf{X},\mathbf{\Omega})=\sum_{i=1}^n L(\mathbf{X}_i,\mathbf{\Omega})/n=L(\mathbf{S},\mathbf{\Omega})$ and use $PL(\mathbf{X},\mathbf{\Omega})$ to denote the corresponding population mean for a fixed precision matrix $\mathbf{\Omega}$. Let $G_n=\sqrt{n}(P_n-P)$ be the empirical process.
We write
\begin{eqnarray*}
&&L(\mathbf{S},\hat{\mathbf{\Omega}}(\lambda))-L(\mathbf{S},\hat{\mathbf{\Omega}}(\lambda_n))\\
&\le&P_nL(\mathbf{X},\hat{\mathbf{\Omega}}_{\mathcal{S}_\lambda})-P_nL(\mathbf{X},\hat{\mathbf{\Omega}}(\lambda_n))\\
&=&P_nL(\mathbf{X},\hat{\mathbf{\Omega}}_{\mathcal{S}_\lambda})-PL(\mathbf{X},\hat{\mathbf{\Omega}}_{\mathcal{S}_\lambda})-\left(P_nL(\mathbf{X},\hat{\mathbf{\Omega}}(\lambda_n))-PL(\mathbf{X},\hat{\mathbf{\Omega}}(\lambda_n))\right)\\
&&+PL(\mathbf{X},\hat{\mathbf{\Omega}}_{\mathcal{S}_\lambda})-PL(\mathbf{X},\mathbf{\Omega}_0)-\left(PL(\mathbf{X},\hat{\mathbf{\Omega}}(\lambda_n))-PL(\mathbf{X},\mathbf{\Omega}_0)\right),
\end{eqnarray*}
 where $\mathbf{\Omega}_0$ is the true precision matrix.
Define $M_1(\mathbf{X})=\sqrt{n}\left(L(\mathbf{X},\hat{\mathbf{\Omega}}_{\mathcal{S}_\lambda})-L(\mathbf{X},\mathbf{\Omega}_0)\right)$, and $M_2(\mathbf{X})=\sqrt{n}\left(L(\mathbf{X},\hat{\mathbf{\Omega}}(\lambda_n))-L(\mathbf{X},\mathbf{\Omega}_0)\right)$, then the previous display can be written as
\begin{eqnarray*}
&&L(\mathbf{S},\hat{\mathbf{\Omega}}(\lambda))-L(\mathbf{S},\hat{\mathbf{\Omega}}(\lambda_n))\\
&\le&\frac{1}{n}G_n M_1(\mathbf{X})-\frac{1}{n}G_n M_2(\mathbf{X})\\
&&+\left(PL(\mathbf{X},\hat{\mathbf{\Omega}}_{\mathcal{S}_\lambda})-PL(\mathbf{X},\mathbf{\Omega}_0)\right)-\left(PL(\mathbf{X},\hat{\mathbf{\Omega}}(\lambda_n))-PL(\mathbf{X},\mathbf{\Omega}_0)\right)\\
&=:&\mbox{(I)+(II)+(III)+(IV)}.
\end{eqnarray*}
We first bound (I) from above. Classical maximum likelihood theory tells us that $\mathbf{H}_n:=\sqrt{n}(\hat{\mathbf{\Omega}}_{\mathcal{S}_\lambda}-\mathbf{\Omega}_0)$ is asymptotically normal (for the non-constrained elements of the matrix). Applying Lemma 19.31 in \cite{vaart98}, we have $G_n(M_1(\mathbf{X})-\mathbf{H}_n^T dL(\mathbf{X},\mathbf{\Omega}_0)/d\mathbf{\Omega})\stackrel{P}{\rightarrow}0$ and then it is easily seen that $G_n(M_1(\mathbf{X}))=O_P(1)$. Similarly, given that \cite{fan08} have shown that $\hat{\mathbf{\Omega}}(\lambda_n)$ is also asymptotically normal, we have $G_nM_2(\mathbf{X})=O_P(1)$. For (III) and (IV), a simple Taylor expansion around $\mathbf{\Omega}$ yields both term to be of order $O_P(1/n)$.

\section*{ACKNOWLEDGEMENT}
This research is supported by Singapore Ministry of Education Tier 1 SUG.

\bibliographystyle{asa}
\bibliography{papers.txt,books.txt}
\end{document}